\documentclass{elsart}
\usepackage{graphics}
\usepackage{amssymb}
\journal{Solid State Communications}
\begin{document}

\begin{frontmatter}

\title{Magnetospectroscopy of epitaxial few-layer graphene}

\author[label1]{M.L. Sadowski}
\author[label1]{G. Martinez},
\author[label1]{M. Potemski},
\author[label2,label3]{C. Berger},
\author[label2]{W.A. de Heer}

\address[label1]{Grenoble High Magnetic Field Laboratory, Grenoble,France}
\address[label2]{Georgia Institute of Technology, Atlanta, Georgia, USA}
\address[label3]{Institut N\'{e}el, CNRS, Grenoble, France}

\begin{abstract}
The inter-Landau level transitions observed in far-infrared
transmission experiments on few-layer graphene samples show a
behaviour characteristic of the linear dispersion expected in
graphene. This behaviour persists in relatively thick samples, and
is qualitatively different from that of thin samples of bulk
graphite.
\end{abstract}

\end{frontmatter}


The interest in two-dimensional graphite is fuelled by its
particular band structure and ensuing dispersion relation for
electrons, leading to numerous differences with respect to
``conventional" two-dimensional electron gases (2DEG). Single
graphite layers (graphene) have long been used as a starting point
in band structure calculations of bulk graphite
\cite{{Wallace},{McClure56},{Slonczewski}} and, more recently,
carbon nanotubes \cite{AndoCNTreview}. The band structure of a
single graphene sheet is considered to be composed of cones
located at two inequivalent Brillouin zone corners at which the
conduction and valence bands merge. In the vicinity of these
points the electron energy depends linearly on its momentum, which
implies that free charge carriers in graphene are governed not by
Schr\"{o}dinger's equation, but rather by Dirac's relativistic
equation for zero rest mass particles, with an effective velocity
$\tilde{c}$, which replaces the speed of light
\cite{{Haldane},{AndoPR02}}.

The recent appearance of ultrathin graphite layers (few-layer
graphene, FLG), obtained by epitaxial
\cite{{Forbeaux},{Charrier},{Berger1}} and exfoliation techniques
\cite{NovoselovSci}, followed by single graphene and its unusual
sequence of quantum Hall states \cite{{NovoselovNat},{ZhangNat}}
has re-ignited this interest. The prospects of studying quantum
electrodynamics in solid state experiments on the one hand and the
possibility of future applications in carbon-based electronics on
the other are currently driving a considerable research effort.
The majority of the published literature remains theoretical; the
extremely small lateral dimensions ($\approx 10\mu$m) of the
graphene flakes used in the above-mentioned transport experiments
makes them difficult objects for experimental studies. Moreover,
due to the somewhat hit-and-miss character of the exfoliation
method, as well as the inherent difficulty of obtaining large
numbers of samples, it appears to be an unlikely candidate for
possible applications. Epitaxial methods on the other hand offer
the opportunity of obtaining relatively large, high quality
two-dimensional graphite \cite{Berger2}. In the following, we
present optical measurements of the characteristic dispersion
relation of FLG, confirming directly its linear (``relativistic")
character.

A number of epitaxial graphene samples have been studied by means
of far-infrared magnetotransmission measurements. The samples were
about 4 $\times$ 4 mm$^2$ in area, grown by sublimating SiC
substrates at high temperatures \cite{Berger1, Berger2}. The
experimental details and part of the results have been described
elsewhere \cite{Sadowski}.

\begin{figure}[bt]
\begin{center}
\includegraphics{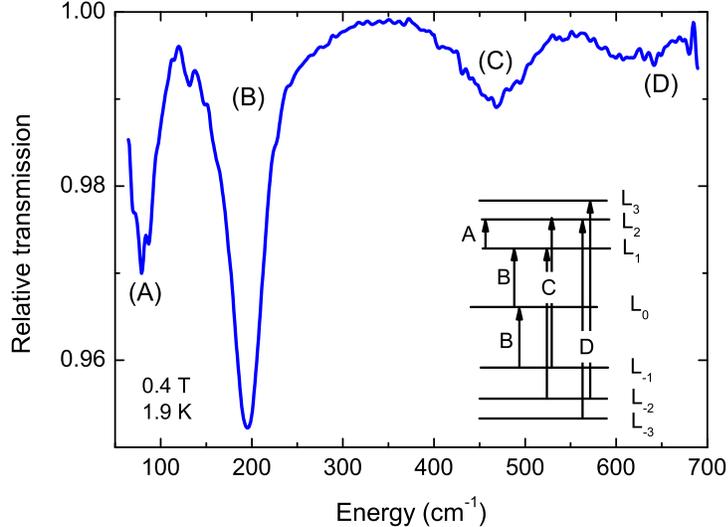}
\caption{Transmission spectrum of epitaxial graphene at 0.4 T. The
inset shows a schematic of the assignations of the observed
transitions.} \label{spectrum_figure}
\end{center}
\vspace*{8pt}
\end{figure}

\begin{figure}
\begin{center}
\includegraphics{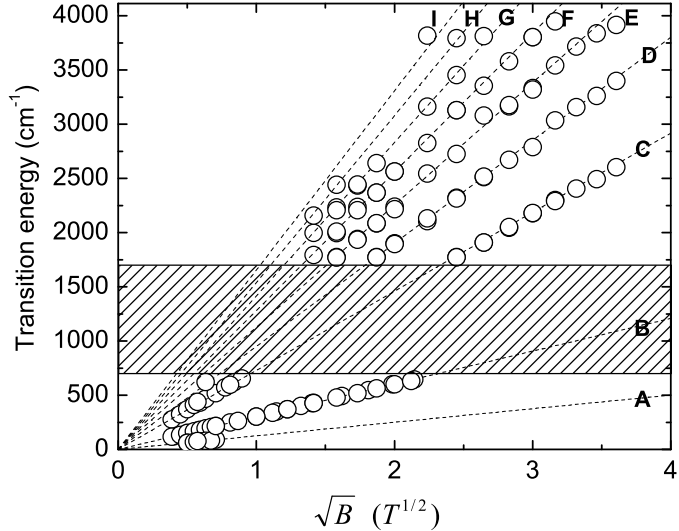}
\caption{Evolution with magnetic field of transitions observed in
transmission. The letters correspond to those used in
Fig.~\ref{spectrum_figure}; the shaded region corresponds to the
range where the substrate is opaque.}\label{positions_figure}
\end{center}
\vspace*{8pt}
\end{figure}

A representative transmission spectrum of a three-graphene-layer
sample is shown in Fig.~\ref{spectrum_figure} for a weak magnetic
field of 0.4 T. When the magnetic field is increased, all the
features visible in this figure are displaced towards higher
energies. Furthermore, their strength increases \cite{Sadowski}
and more features become visible at higher energies. The positions
of the features observed for the sample containing 3 graphene
layers are plotted versus the square root of the magnetic field in
Fig.~\ref{positions_figure}. It may be seen that the resonant
energies observed evolve proportionally to the square root of the
magnetic field. The oscillator strength of the transition labelled
B in Fig~\ref{spectrum_figure} has also been shown \cite{Sadowski}
to increase linearly with the square root of the magnetic field.

These results are, in a first approximation, in excellent
agreement with predictions arising from a simple single-particle
model of non-interacting massless Dirac fermions.

Using appropriate graphene wavefunctions \cite{AndoCNTreview} and
the Hamiltonian commonly used to describe electrons in a single
graphene layer, it is fairly straightforward to work out the
optical selection rules \cite{Sadowski2}. It may then be shown
that the allowed transitions are $L_n \rightarrow L_m$ such that
$|m|=|n|-1$ for the ``+" circular polarisation and $|m|=|n|+1$ in
the ``-" circular polarisation. For unpolarised radiation, used in
the current experiment, the allowed transitions are simply those
between states $n,m$ such that $|m|=|n|\pm 1$. The Landau level
energies are obtained as
\begin{equation}
E_n = \tilde{c} \sqrt{2 \hbar e B |n|}
\end{equation}
where $\tilde{c}$ is the effective velocity of the Dirac fermions,
B is the magnetic field and $n = 0, \pm 1, \pm2$ ... is the Landau
level index (the electron and hole levels being identical). The
energies of the allowed optical transitions may then be concisely
written as
\begin{equation}
E_{n}^t = \tilde{c} \sqrt{2 \hbar e B} (\sqrt{|n + 1|}\pm
\sqrt{|n|})
\end{equation}

The positions of the transitions shown in
Fig.~\ref{positions_figure} are summarised in Table~\ref{tab_a}.

\begin{table}
\begin{center}
\begin{tabular}{lcrr}
\hline {Line} & {Slope in units of $\tilde{c}\sqrt{2e\hbar}$}
& {Transition}   \\
\hline
 A & $\sqrt{2}-\sqrt{1}$& $L_1
\rightarrow L_2$\\
  B & 1 & $L_0 \rightarrow L_1 (L_{-1} \rightarrow L_0)$\\
  C & $\sqrt{2}+\sqrt{1}$ & $L_{-1} \rightarrow L_2 (L_{-2} \rightarrow
L_1)$\\
  D & $\sqrt{3}+\sqrt{2}$ & $L_{-2} \rightarrow L_3 (L_{-3}
\rightarrow L_2)$\\
  E & $\sqrt{4}+\sqrt{3}$ & $L_{-3} \rightarrow L_4 (L_{-4}
\rightarrow L_3)$\\
  F & $\sqrt{5}+\sqrt{4}$ & $L_{-4} \rightarrow L_5 (L_{-5}
\rightarrow L_4)$\\
  G & $\sqrt{6}+\sqrt{5}$ & $L_{-5} \rightarrow L_6 (L_{-6}
\rightarrow L_5)$\\
  H & $\sqrt{7}+\sqrt{6}$ & $L_{-6} \rightarrow L_7 (L_{-7}
\rightarrow L_6)$\\
  I & $\sqrt{8}+\sqrt{7}$ & $L_{-7} \rightarrow L_8 (L_{-8}
\rightarrow L_7)$\\
 \hline
\end{tabular}
\caption{Observed lines and their assignments} \label{tab_a}
\end{center}
\end{table}

It should be stressed that all the positions of all the observed
lines are described by a single fitting parameter - the effective
light velocity $\tilde{c}$. We should add, for the sake of
completeness, that the present experiment, using unpolarised
light, does not distinguish between electrons and holes, which are
expected to be identical in terms of the effective mass and
dispersion relation. Thus, transition A, attributed to the $L_1
\rightarrow L_2$ process, could also be due to the corresponding
$L_{-2} \rightarrow L_{-1}$ one. While a p-type character appears
to be unlikely, it cannot be ruled out on the basis of the
experiment in question.

The striking agreement of the experimental data obtained using
several layers of graphene with expectations for a single layer is
surprising, given that calculations suggest a completely different
behaviour already for a graphene bilayer \cite{Abergel}. On the
other hand, it has long been known that particles with a linear
dispersion exist in bulk graphite as well - a minority pocket of
carriers in the vicinity of the H point of the Brillouin zone were
shown to give rise to electronic transitions following a square
root dependence on the magnetic field \cite{Toy}. The question
therefore is posed: at what point, if at all, does epitaxial FLG
become bulk graphite?

Early work on epitaxial graphene \cite{Forbeaux} suggested that
the process of baking SiC substrates led to a single graphene
layer floating above a graphite layer. More recent calculations
\cite{Varchon} suggest that the first carbon layer on top of an
SiC substrate has an electronic structure different from that of
graphene, and acts as a buffer, allowing subsequent layers to
behave like graphene.  A strong dependence of the electronic
structure of FLG on the type of stacking has also been suggested
\cite{Guinea}. The common Bernal, or AB, stacking found for
example in HOPG graphite is usually assumed for all FLG structures
as well; this is not necessarily the case. Also, let us note that
the HOPG interlayer distance of 3.354 $\AA$ may not be the correct
value for epitaxial graphene.

\begin{figure}
\begin{center}
\includegraphics{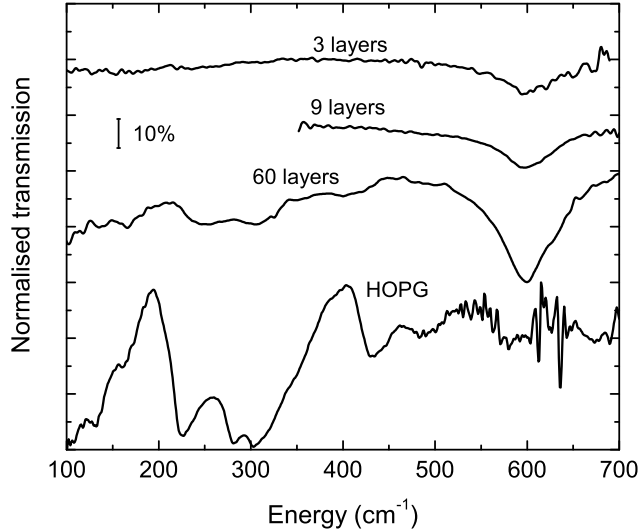}
\caption{Transmission spectra at 4T for epitaxial FLG samples (top
three) of varying thickness and, for comparison, of HOPG graphite
at the same magnetic field.}\label{samples_spectra_figure}
\end{center}
\vspace*{8pt}
\end{figure}

In order to elucidate the effect of multiplying layers on the
transmission spectrum, samples of varying thickness were studied
and compared with a layer of HOPG obtained by exfoliation. The
details of this study shall be presented elsewhere
\cite{Sadowski3}; for the time being let us note the qualitative
differences in the spectra, shown in
Fig.~\ref{samples_spectra_figure}. Four spectra are shown, at a
magnetic field of 4T: for sample consisting of ~3, ~9 and ~60
layers of graphene on SiC, and for the HOPG sample. The dominant
feature in the epitaxial samples is always the $L_0 \rightarrow
L_1$ ($L_{-1} \rightarrow L_0$) transition; we can see that it
grows stronger as the number of layers is increased, and is
several times stronger for the sample containing 60 layers. In
this sample one can also see the appearance of other features at
lower energies, which were absent in the thinner samples, and
which appear to correspond to bulk-like features visible in the
lowest (HOPG) trace in the figure. On the other hand, the $L_0
\rightarrow L_1$ ($L_{-1} \rightarrow L_0$) transition, which has
a square root dispersion even in the 60 layer sample, is absent
from the HOPG spectrum.

The observed persistence of the Dirac fermion-like behaviour of
the carriers in epitaxial FLG up to relatively thick (~19 nm)
structures appears to suggest that the structure of this material
is in fact different from that of bulk HOPG. The simplest
explanation would be a far weaker interaction between adjacent
graphene layers, leading to a sequence of graphene layers instead
of bulk, or even multilayer, graphene. More studies are necessary
to elucidate this question.

The GHMFL is a ``Laboratoire conventionn\'{e} avec l'UJF et l'INPG
de Grenoble". The present work was supported in part by the
European Commission through grant RITA-CT-2003-505474 and by
grants from the Intel Research Corporation and the NSF: NIRT
``Electronic Devices from Nano-Patterned Epitaxial Graphite".

\end{document}